\documentclass[preprint]{elsarticle}
\usepackage{amsfonts}
\usepackage{amsmath}
\usepackage{amssymb}
\usepackage{graphicx}
\usepackage{color}
\usepackage{aas_macros}
\begin{document}

\title{Bose-Einstein condensation in relativistic plasma}

\author[M,U]{M. A. Prakapenia\corref{cor1}}
\ead{nikprokopenya@gmail.com}

\author[M,P,R]{G. V. Vereshchagin}
\ead{veresh@icra.it}

\cortext[cor1]{Corresponding author}

\address[M]{ICRANet-Minsk, Institute of physics, National academy of sciences of Belarus\\ 220072 Nezale\v znasci Av. 68-2, Minsk, Belarus}
\address[U]{Department of Theoretical Physics and Astrophysics, Belarusian State University\\ 220030 Nezale\v znasci Av. 4, Minsk, Belarus}
\address[P]{ICRANet, 65122 Piazza della Repubblica, 10, Pescara, Italia}
\address[R]{INAF -- Istituto di Astrofisica e Planetologia Spaziali, 00133 Via del
Fosso del Cavaliere, 100, Rome, Italy}


\begin{abstract}
The phenomenon of Bose-Einstein condensation is traditionally associated with
and experimentally verified for low temperatures: either of nano-Kelvin scale
for alkali atoms
\cite{1995Sci...269..198A,1995PhRvL..75.1687B,1995PhRvL..75.3969D} or room
temperatures for quasi-particles
\cite{2002Sci...298..199D,2006Natur.443..430D} or photons in two dimensions
\cite{2010Natur.468..545K}. Here we demonstrate out of first principles that
for certain initial conditions non-equilibrium plasma at relativistic
temperatures of billions of Kelvin undergoes condensation, predicted by
Zeldovich and Levich in their seminal work \cite{1969JETP...28.1287Z}. We
determine the necessary conditions for the onset of condensation and discuss
the possibilities to observe such a phenomenon in laboratory and astrophysical conditions.
\end{abstract}

\maketitle

\section{Introduction}

The phenomenon of quantum condensation of bosons was predicted by Bose and
Einstein \cite{1924ZPhy...26..178B,Ein1,Ein2}. In physics textbooks it is
associated with cooling to low temperatures
\cite{1980stph.book.....L,2008bcdg.book.....P}, and it was indeed observed for
ultracold atoms
\cite{1995Sci...269..198A,1995PhRvL..75.1687B,1995PhRvL..75.3969D} and for
quasi-particles \cite{2002Sci...298..199D,2006Natur.443..430D}. It was also
observed for photons in a microcavity \cite{2010Natur.468..545K}, where
special boundary conditions ensured necessary pre-requisites for Bose-Einstein
condensation: (a) photon number conservation or (b) generation of effective
mass via spatial confinement \cite{2014JPhB...47x3001K}.

Following the original prediction, Bose-Einstein condensation (BEC) is
understood as a quantum phenomenon, occurring in ideal Bose gas of massive
particles when the temperature decreases at constant density, or,
alternatively, when the density of particles increases at fixed temperature,
leading to condensation of a fraction of particles in the lowest energy state. It consists of appearance of a separate phase in a gas whose particles occupy the lowest quantum state. This phenomenon is traditionally associated with
low temperatures, as well demonstrated by cooling alkali atoms to nanokelvin degrees \cite{1995Sci...269..198A,1995PhRvL..75.1687B,1995PhRvL..75.3969D}.

In a pure photon gas such phenomenon cannot occur, because photons are
massless particles and cooling leads to disappearance of photons. Nevertheless, it was predicted in opaque plasma in a
pioneering work \cite{1969JETP...28.1287Z}. There it was conjectured that, in
absence of photon absorption, the dominant interaction process in a rarefied
hot plasma is Compton scattering, which conserves photon number, fulfilling
the condition (a). Considering the properties of nonlinear Kompaneets equation
\cite{1956JETP...04..730K}, which describes time evolution of photon spectrum
due to Compton scattering on nonrelativistic electrons with Maxwellian
distribution, the mechanism of condensation was illustrated. It was shown
that, unlike ideal Bose gases, BEC manifests itself as an excess of photons over the Planck
distribution \cite{1986PhyA..139..165M}, which is only possible at intermediate
energies: between the spectral peak and the critical energy, below which
absorption dominates. It was proposed that such phenomenon may occur in
astrophysical conditions, when hot radiation passes through cold plasma. A
related phenomenon called comptonization is indeed observed, albeit in an
almost transparent plasma, and is known as Sunyaev-Zeldovich effect
\cite{1969Ap&SS...4..301Z,1970Ap&SS...7....3S}. The photon condensation in
plasma is a bulk phenomenon, arising in homogeneous and unbounded system;
moreover, photons in plasma acquire effective mass \cite{2017PhRvA..95f3611M}
ensuring condition (b). The mechanism of BEC of photons in thermodynamic
equilibrium with the atoms of diluted gases has been also discussed in
\cite{2013PhRvA..88a3615K}, while in \cite{2019arXiv190207998M}\ statistical
theory of photon condensation is developed.

The purpose of the present work is to demonstrate out of first principles that
for certain initial conditions photons in non-equilibrium optically thick
electron-positron plasma undergo BEC. Thus, in contrast with the traditional
belief that BEC is related to cooling to low temperatures, we provide an
example for photon condensation at very high, relativistic temperatures. We
show, that although the condition (a) is violated, as the number of particles
changes due to triple interactions, BEC is present as a transient phenomenon
both at nonrelativistic and relativistic temperatures.


\section{Kinetic versus thermal equilibrium}

Considering relaxation of non-equilibrium electron-positron-photon plasma
\cite{2007PhRvL..99l5003A}, as well as relativistic (average energy per
particle exceeds its rest mass energy)\ plasma with proton load
\cite{2009PhRvD..79d3008A} with arbitrary initial conditions it was found that
this process occurs in two steps. First, detailed balance is established in
two-particle (binary) interactions, such as Compton and Coulomb scattering,
pair creation and annihilation in two photons. This balance corresponds to a
metastable state called kinetic equilibrium, which is characterized by the
same temperature $T_{k}$\ of all particles, and nonzero chemical potentials
$\mu_{i}$. It is important to stress that condition (a) is satisfied in kinetic equilibrium. The distribution function of particles with energy $E$ in this
state has the form%
\begin{equation}
f=\frac{2}{\left(  2\pi\hbar\right)  ^{3}}\frac{1}{\exp\left(  \frac{E-\mu
_{i}}{kT_{k}}\right)  \pm1},\label{keq}%
\end{equation}
where $\hbar$ is reduced Planck's constant $k$\ is Boltzmann's constant, and
signs "$+$" and "$-$" correspond to Fermi-Dirac and Bose-Einstein statistics,
respectively. Kinetic equilibrium is established if the rates of binary
interactions exceed the ones of three-particle (triple) interactions, namely
relativistic bremsstrahlung, double Compton scattering, radiative pair
production and three-photon annihilation. The characteristic timescale of kinetic equilibrium can be estimated as 
\begin{equation}
t_{k}\simeq A\left(  \sigma_{T}%
n_{e}c\right)  ^{-1}, \label{tKEQ}
\end{equation}
where $\sigma_{T}$ is the Thomson cross section,
$n_{e}$\ is electron number density, $c$\ is the speed of light, and the coefficient $A\simeq 20$ \cite{2007PhRvL..99l5003A}. When triple interactions finally come into detailed balance, thermal equilibrium is
established and the chemical potential of photons vanishes.

In a recent work \cite{2019PhLA..383..306P} quantum statistics of particle was
accounted for, and in addition the rates of triple interactions were evaluated
directly from the QED matrix elements. It was found that kinetic equilibrium
is established in the relaxation process prior to the thermal one only for
nonrelativistic plasma, with final thermal equilibrium temperatures
$T_{th}<0.3m_{e}c^{2}/k$.

The possibility of condensation of photons in kinetic equilibrium state occurs
if the initial number density of photons $n_{\gamma}$\ exceeds the one given
by eq. (\ref{keq}) with zero chemical potential of photons $\mu_{\gamma}=0$,
namely%
\begin{equation}
n_{\gamma}>\frac{2\zeta(3)}{\pi^{2}}\left(  \frac{\hbar}{mc}\right)
^{-3}\left(  \frac{kT_{k}}{m_{e}c^{2}}\right)  ^{3},\label{conddens}%
\end{equation}
where $\zeta\left(  s\right)  $ is the Riemann $\zeta$-function. Provided that
binary interactions do not change the number of photons, these particles tend
to accumulate in lowest energy states and form an excess over the Planck
distribution as long as kinetic equilibrium is maintained. When the rates of
triple interactions become significant, the number of photons reduces and the
excess over the Planck distribution tends to disappear and plasma relaxes
towards thermal equilibrium described by the distribution (\ref{keq}) with
zero chemical potential of photons. The characteristic timescale of thermal
equilibrium can be estimated as $t_{th}\sim\left(  \alpha\sigma_{T}%
n_{e}c\right)  ^{-1}$, where $\alpha$\ is the fine structure constant.
Detailed calculations show that both kinetic and thermal equilibrium
timescales are functions of total energy density
\cite{2010PhRvE..81d6401A,2019PhLA..383..306P} and at high temperatures
$T_{th}>0.3m_{e}c^{2}/k$\ they nearly coincide.

We have found that BEC of photons may occur for a broad class of initial
distribution functions, including Gaussian and Wien
distributions. In what follows we give several examples.

\section{Boltzmann equations}

We solve relativistic Boltzmann equations \cite{2017rkt..book.....V} for
one-particle distribution functions of electrons $e^{-}$, positrons $e^{+}$
and photons $\gamma$ with quantum corrections described by Uehling-Uhlenbeck
collision integrals
\begin{equation}
\frac{1}{c}\frac{\partial f_{i}}{\partial t}=\sum_{q}\left(  \eta_{i}^{q}%
-\chi_{i}^{q}f_{i}\right)  ,\label{BE}%
\end{equation}
where $f_{i}(\epsilon,t)$ are their distribution functions, index $i$ denotes
the sort of particles, $\epsilon$ is their energy, $\eta_{i}^{q}$ and
$\chi_{i}^{q}$ are the emission and the absorption coefficients of a particle
of type "$i$" via the physical process labelled by $q$. All binary and triple
interactions between particles are taken into account. \ The number density
and spectral energy density are defined as follows
\[
n_{i}=\int f_{i}(\mathbf{p},t)d^{3}p;\quad\quad\frac{d\rho_{i}}{d\epsilon
}=\frac{4\pi}{c^{2}}|\mathbf{p}|\epsilon^{2}f_{i}.
\]
The emission and absorption coefficients for the particle $I$ in a binary
process $I+II\leftrightarrow III+IV$ have the following form
\begin{align}
\eta_{I}^{2p} &  =\int d^{3}p_{2}d^{3}p_{3}d^{3}p_{4}\ W_{(3,4|1,2)}%
\ \label{eta2p}\\
&  \times f_{III}f_{IV}\left(  1+\xi f_{I}\right)  \left(  1+\xi
f_{II}\right)  ,\nonumber\\
\chi_{I}^{2p}f_{I} &  =\int d^{3}p_{2}d^{3}p_{3}d^{3}p_{4}\ W_{(1,2|3,4)}%
\label{chi2p}\\
&  \times\ f_{I}f_{II}\left(  1+\xi f_{III}\right)  \left(  1+\xi
f_{IV}\right)  ,\nonumber
\end{align}
where transition rates are $W_{(3,4|1,2)}d^{3}p_{3}d^{3}p_{4}=Vdw_{(3,4|1,2)}$
and $W_{(1,2|3,4)}d^{3}p_{1}d^{3}p_{2}=Vdw_{(1,2|3,4)}$, $V$ is normalization
volume, $dw$ is differential reaction probability per unit time, $\xi=\psi
h^{3}/2$ and $\psi$ is +1,-1 for Bose-Einstein and Fermi-Dirac statistic,
respectively, $h=2\pi\hbar$. The emission and absorption coefficients for the
particle $I$ in a triple process $I+II\leftrightarrow III+IV+V$ have the
following forms
\begin{align}
\eta_{I}^{3p} &  =\int d^{3}p_{2}d^{3}p_{3}d^{3}p_{4}d^{3}p_{5}%
\ W_{(3,4,5|1,2)}\label{eta3p}\\
&  \times\ f_{III}f_{IV}f_{V}\left(  1+\xi f_{I}\right)  \left(  1+\xi
f_{II}\right)  ,\nonumber\\
\chi_{I}^{3p}f_{I} &  =\int d^{3}p_{2}d^{3}p_{3}d^{3}p_{4}d^{3}p_{5}%
\ W_{(1,2|3,4,5)}\label{chi3p}\\
&  \times\ f_{I}f_{II}\left(  1+\xi f_{III}\right)  \left(  1+\xi
f_{IV}\right)  \left(  1+\xi f_{V}\right)  ,\nonumber
\end{align}
where $W_{(3,4,5|1,2)}d^{3}p_{3}d^{3}p_{4}d^{3}p_{5}=Vdw_{(3,4,5|1,2)}$ and
$W_{(1,2|3,4,5)}d^{3}p_{1}d^{3}p_{2}=V^{2}dw_{(1,2|3,4,5)}$. The expression
for $dw$ is given in quantum electrodynamics as
\begin{align}
dw &  =c(2\pi\hbar)^{4}\delta(\epsilon_{in}-\epsilon_{f\!in})\delta
(\mathbf{p}_{in}-\mathbf{p}_{f\!in})|M_{fi}|^{2}V\nonumber\\
&  \times\left(  \prod_{in}\frac{\hbar c}{2\epsilon_{in}V}\right)  \left(
\prod_{f\!in}\frac{d^{3}p_{f\!in}}{(2\pi\hbar)^{3}}\frac{\hbar c}%
{2\epsilon_{f\!in}}\right)  ,\label{mel}%
\end{align}
where $p_{f\!in}$ and $\epsilon_{f\!in}$ are respectively momenta and energies
of outgoing particles, $p_{in}$ and $\epsilon_{in}$ are momenta and energies
of incoming particles, $M_{fi}$ is the corresponding matrix element, $\delta
$-functions stand for energy-momentum conservation. Therefore, collision
integrals, i.e. right-hand side of equations (\ref{BE}), are integrals over
the phase space of interacting particles, which include the quantum
electrodynamics matrix elements, see e.g.
\cite{2017rkt..book.....V,Berestetskii1982} for binary reactions and
\cite{1952RSPSA.215..497M} for double Compton scattering,
\cite{2004epb..book.....H} for relativistic bremsstrahlung and
\cite{1976spr..book.....J} for substitution rules in computation of remaining
matrix elements for triple reactions.

\section{Numerical results}

The coupled system of integro-differential equations (\ref{BE}) is solved
numerically using a finite difference method by introducing a computational
grid in the phase space to represent the distribution functions and to compute
collisional integrals
\cite{2007PhRvL..99l5003A,2018JCoPh.373..533P,2019PhLA..383..306P}. Thus the
system of integro-differential equations is reduced to a system of stiff
ordinary differential equations which are solved by the Gear method
\cite{2017rkt..book.....V,1976oup..book.....H}. Due to a finite numerical
resolution, the effective photon mass in plasma \cite{2017PhRvA..95f3611M}
always turns out to be outside our grid, therefore we consider photons as
massless particles.

Evidently, the possibility of the development of photon condensation strongly
depends on the initial state of plasma. An example considered in
\cite{1969JETP...28.1287Z} is the initial state with hot photons and cold
electrons, where photons have the Planck spectrum with temperature $T_{\gamma
}$, and electrons have the Maxwellian distribution with temperature
$T_{e}<T_{\gamma}$. It is expected that Compton scattering redistributes
energy between these components so that photons would cool down. Given that
the number of particles is conserved, photons are expected to undergo condensation.

It turns out that these initial conditions are not suitable to obtain BEC. In
our simulations with cold degenerate electrons and high temperature photons
during the relaxation process the cooling of photons is accompanied with the
decrease of their number density, due to triple reactions not properly taken
into account in \cite{1969JETP...28.1287Z} in such a way that their
condensation does not appear \cite{Prakapenia2019}. Besides, at relativistic
temperatures annihilation of photons is efficient enough and prevents accumulation of
photons at low energies. Therefore, the necessary condition for BEC of photons
cannot be met with such initial conditions.

We found that in order to favour BEC, initial distribution
of photons should not be broader than Wien spectrum, and the peak of the
distribution should be above the critical energy, below which triple
interactions dominate over the binary ones. Initial state for pairs can be
arbitrary, and for high temperature plasma pairs initially can be even absent:
they are quickly produced from photons. As Coulomb interactions are much
faster than Compton scattering and two-photon creation/annihilation
\cite{2009PhRvD..79d3008A}, initial distribution function of pairs acquires
the Fermi-Dirac form well before balance is achieved for Compton scattering
and two-photon creation/annihilation, so photons interact essentially with
Maxwellian electrons.

\begin{figure}[pth]
\centering
\includegraphics[width=\columnwidth]{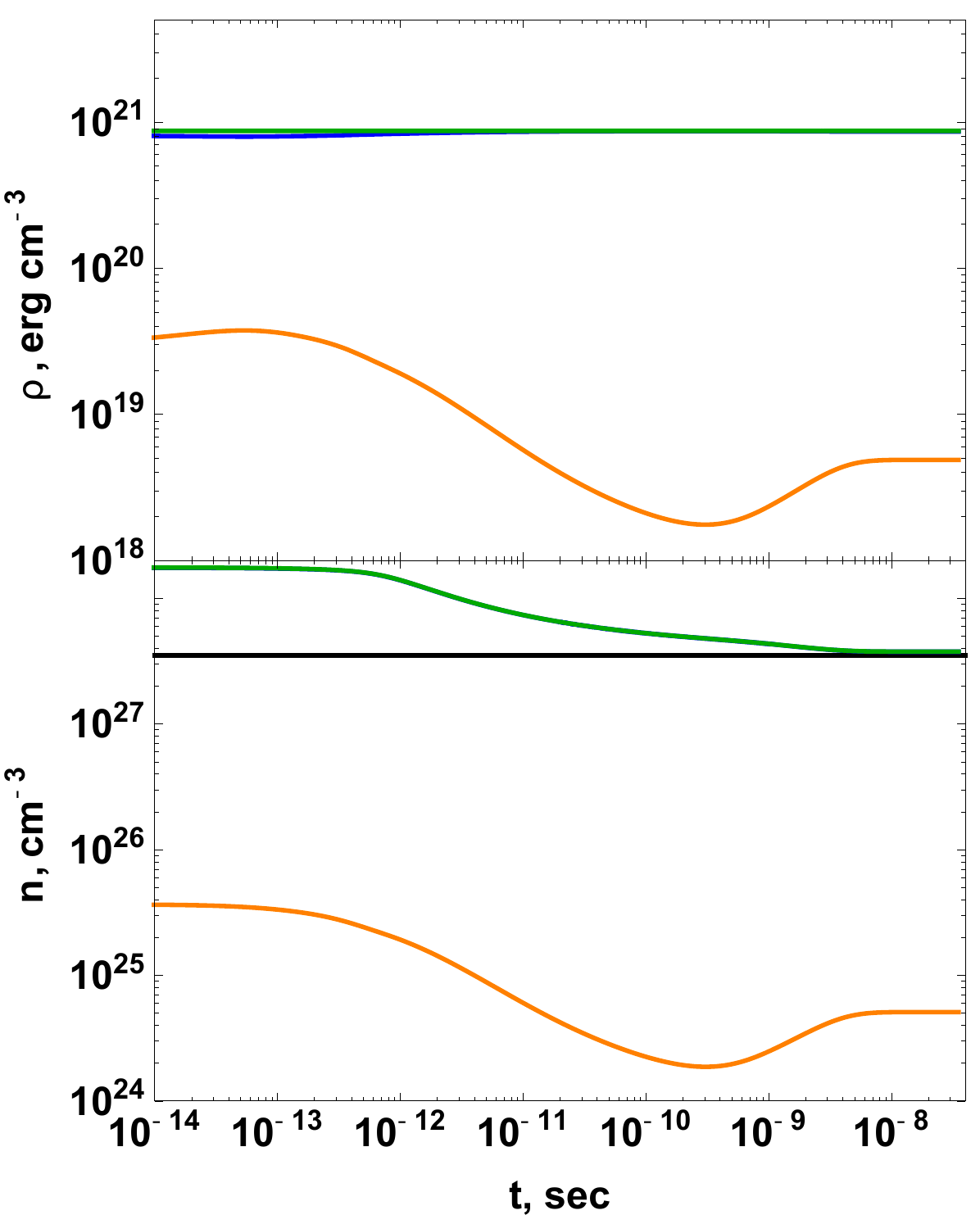}\caption{Time evolution of energy
density (top) and particle number density (bottom) of photons (blue),
electrons/positrons (orange), all together (green) in nonrelativistic case.
Black line represents the final equilibrium quantity. Final equilibrium
temperature is $\theta=k_{B}T/m_{e}c^{2}\simeq0.1$.}%
\label{fig:ron1}%
\end{figure}

\begin{figure}[pth]
\centering
\includegraphics[width=\columnwidth]{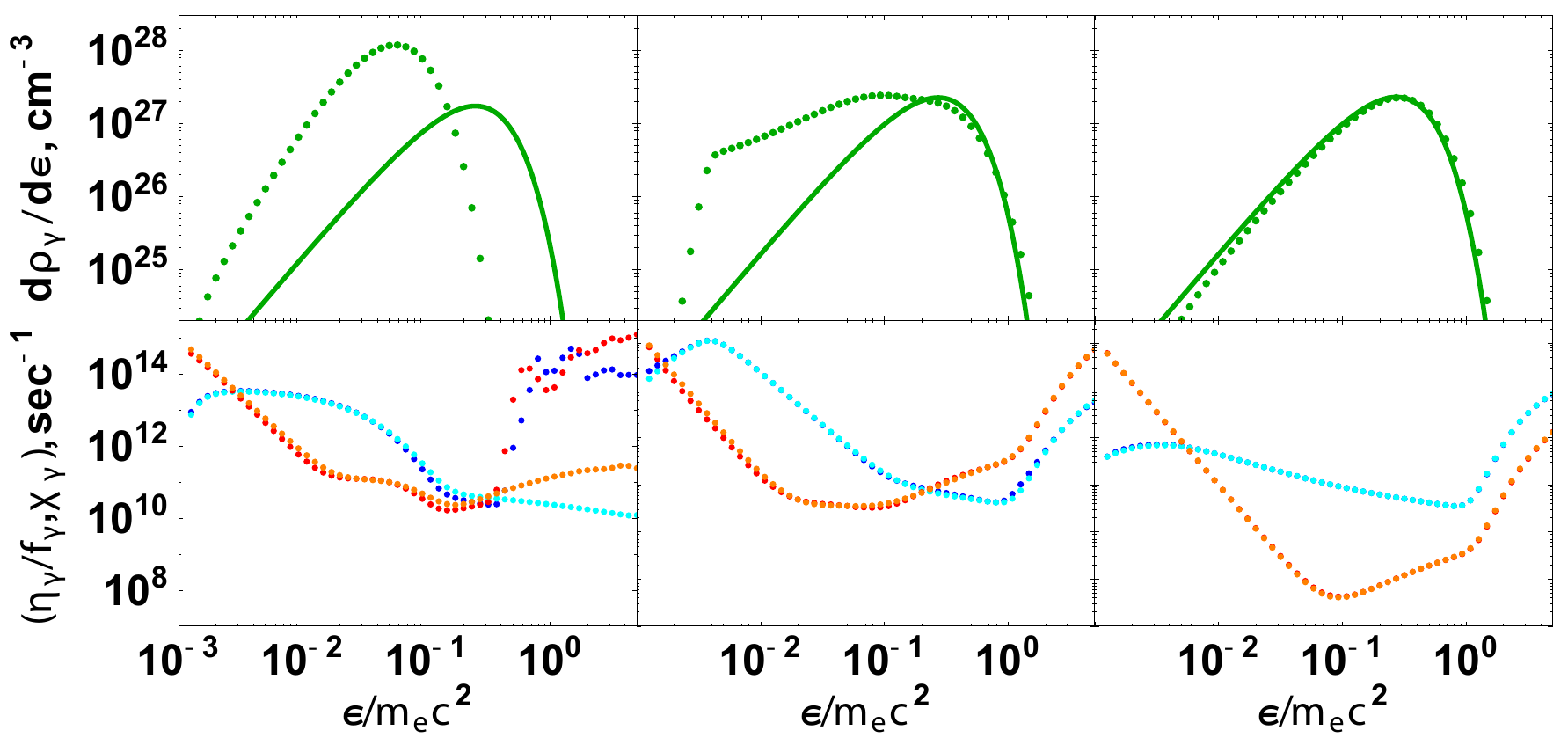}\caption{Top: The spectral
energy density (dots) with the associated Planck fit (solid) for selected time
moments from left to right: $10^{-15},10^{-11},10^{-8}$ sec, in
nonrelativistic case. Bottom: emission and absorption coefficients for photons
(binary reactions: emission (blue) and absorption (cyan); triple reactions:
emission (purple) and absorption (red). The left panel represents initial
distribution of photons; the middle one shows photon condensation, while the
right one corresponds to the final state.}%
\label{fig:sp_rates1}%
\end{figure}

Our results show that photon condensation appears as a transient state both in
non-relativistic and relativistic plasma, yet each case has its own
peculiarity. In the non-relativistic case, we present a particular result with
total energy density $\rho_{tot}=8.7\times10^{20}\,\text{erg cm}^{-3}$
corresponding to a final equilibrium temperature $\theta=k_{B}T/m_{e}%
c^{2}=0.098$. Total initial particle number density is $n_{tot}^{in}%
=5n_{tot}^{f\!in}$, where $n_{tot}^{f\!in}=3.5\times10^{27}\,\text{cm}^{-3}$
is the final total particle number density in thermal equilibrium. The initial
energy density spectrum is a Wien distribution centered at the energy
$\varepsilon=0.06m_{e}c^{2}$ (the 27th energy grid node) for photons and a
single peak (delta function) for pairs located at the energy $\varepsilon
=0.001m_{e}c^{2}$ (the 1st energy grid node). Both energy density and particle
number density of pairs are much smaller that that of photons. The time
evolution of energy density and particle number density is presented on
Fig.~\ref{fig:ron1}, while Fig.~\ref{fig:sp_rates1} shows spectral energy
density and emission and absorption coefficients (reaction rates) for selected
time moments. The energy conservation ensures that the total energy density does
not change with time. The total particle number density changes only due to
the imbalance in triple processes, e.g. bremsstrahlung. Before the time moment
$t\simeq10^{-12}$ sec the triple processes play no significant role. Both
photon number and pair number are almost unchanged; photon annihilation is
also not essential (most photons have energy less than $m_{e}c^{2}$). The
photon energy density slightly decreased and the pair energy density slightly
increased due to Compton scattering. After the time moment $t\simeq10^{-12}$
sec the total particle number starts to decrease. As a result at the time
moment $t\simeq10^{-11}$ sec the total particle number density is
$n_{tot}\simeq n_{\gamma}\simeq6.5\times10^{27}\text{cm}^{-3}\simeq
1.85\ n_{tot}^{f\!in}$. Binary processes are balanced in a broad energy region
implying kinetic equilibrium is established and photons are described by the
Bose-Einstein distribution function (\ref{keq}). The maximum number of photons
supported by the equilibrium with pairs given by eq. (\ref{conddens}) is
$3.8\times10^{27}\text{cm}^{-3}$. The photon number density at this moment is
as large as $6.5\times10^{27}$ cm$^{-3}$ implying that the excess of photons
with the density $2.7\times10^{27}\text{cm}^{-3}$, comparable to the number
density of non-condensed photons, should form a condensate. This excess is
indeed visible in the middle panel of Fig. \ref{fig:sp_rates1}.
At low energies bremsstrahlung and double Compton scattering are more
efficient than single Compton scattering, which results in a steep decrease of
the spectrum. Small deviations from the Planck spectrum seen in the right
panel of Fig. \ref{fig:sp_rates1} in the final state are within the numerical
accuracy obtained on the grid (logarithmic and homogeneous, correspondingly)
with 60 intervals in energy and 24 intervals in angles used for the
computation.
As long as triple interactions are not balanced, thermal equilibrium is not
achieved until about $\sim10^{-8}$ sec when the condensation disappears
completely. It implies the condensate is sufficiently long lived state, in
comparison with the kinetic equilibrium time $t_{k}$. Similar results were obtained with different initial conditions, in particular for Gaussian distribution of photons $d\rho/d\epsilon=a \epsilon^3 exp\{-(\epsilon-b)^2/c^2\}$ with $a=10^{48}~\text{erg}^{-3}\text{cm}^{-3}$, $b=3\times10^{-7}~\text{erg}$, $c=6\times10^{-8}~\text{erg}$, implying that the phenomenon is
quite generic.

We also studied the development of BEC for different degree of initial
degeneracy, by increasing the initial photon number density, which exceeded
equilibrium value by a factor 3, 5, 7 and 10. We found that the system loose
memory of initial distribution at the moment when number density starts to
change due to triple interactions, at $10^{-11},10^{-12},5\times
10^{-13},2\times10^{-13}$ sec, correspondingly. At this moment, which we call
pre-condensation, the photon distribution functions still have non-equilibrium
shape, but power law excess over Planck spectrum starts to appear. In all
cases the condensation (with Planck spectrum and the corresponding excess at
intermediate energies) occurs at $10^{-11}$ sec, and thermalization finishes
at $10^{-8}$ sec, independent on the degree of degeneracy.

Finally, we present the results for the relativistic case with a total energy
density $\rho=2.1\times10^{27}$ erg/cm$^{3}$, corresponding to the final
equilibrium temperature $\theta=3$. The initial state is pairless and photons
are placed at the energy node with $\varepsilon=1.06m_{e}c^{2}$.
\begin{figure}[pth]
\centering
\includegraphics[width=\columnwidth]{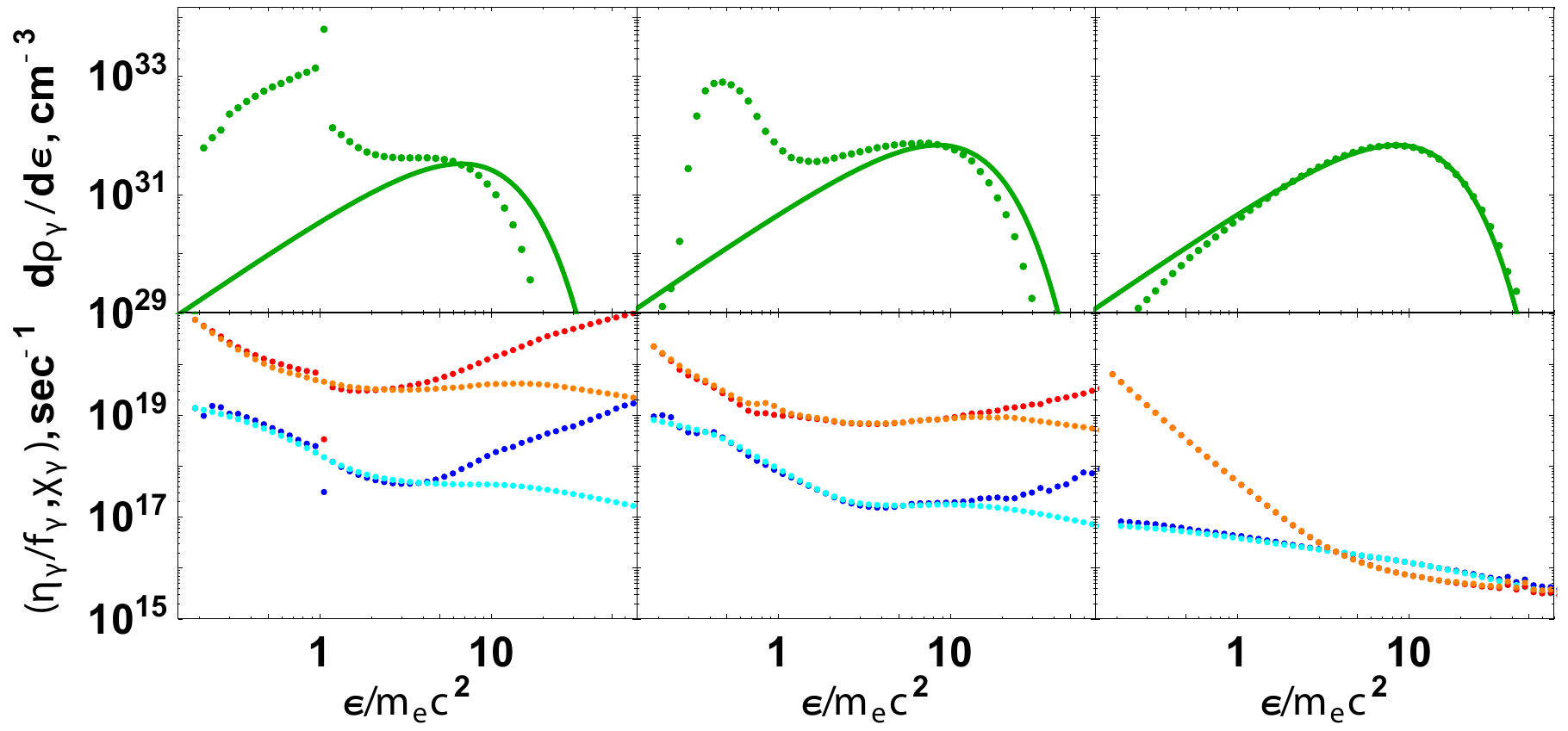}\caption{The spectral
energy density and reaction rates for relativistic case with an initial
degeneracy $n_{tot}^{in}=8.5n_{tot}^{f\!in}$ and final equilibrium temperature
$\theta=3$ at selected time moments (from left to right): $3.8\times
10^{-20},5\times10^{-20},10^{-15}$ sec; colors are as in Fig.
\ref{fig:sp_rates1}.}%
\label{fig:sp_rates_high8}%
\end{figure}Like in nonrelativistic case, BEC develops as a transient state
before the final thermal equilibrium is established. However, unlike the
nonrelativistic case, the photon spectrum does not show the characteristic
power law, but has an excess over the associated Planck spectrum in a form of
a bump. This case is interesting, since it is known \cite{2019PhLA..383..306P}%
\ that triple interactions in plasma with relativistic temperature are faster
than binary ones, see Fig. \ref{fig:sp_rates_high8}. Yet, condensation occurs,
as triple interactions are not fast enough to remove photons in excess over
the Planck spectrum by the time when spectrum acquires an equilibrium shape.

Note that BEC in a microcavity \cite{2010Natur.468..545K}\ represents an
optical analogy of photon condensation discussed in this work. Indeed, photon
number in a microcavity is conserved and after photon thermalization via
interaction with a solution, photons in excess over the thermal number density
undergo condensation. We also point out that in experiments with microcavity
injected photons initially have a narrow energy distribution, similar to the
conditions found in our work. Moreover,\ there is an absorption of photons by
the walls of the cavity, so that photon condensate also represents a transient phenomenon.

\section{Conclusions}

In this work we presented the results of the first principles calculations
demonstrating BEC of photons in relativistic plasma. This phenomenon was
predicted in 1969 and still awaits confirmation in the laboratory.

It is found that condensation of photons may occur on a timescale given by eq. (\ref{tKEQ}) both in nonrelativistic and
in relativistic cases and it manifests in photon spectra described by the
Planck law with an excess formed in the energy range above the critical energy
for the dominance of triple interactions and below the peak of the spectrum.
At nonrelativistic temperatures it is well described by a power law, while at relativistic temperatures it represents a bump. In our nonrelativistic example with $k_{B}T\simeq0.1m_{e}c^{2}$\ the condensation
persist until about $\sim10^{-8}$ sec, when complete thermal equilibrium is established.

It is found that necessary condition for the development of BEC is an excess
of photon number over the equilibrium number, see eq. (\ref{conddens}), as
well as {\bf initial distribution of photons not broader than
Wien spectrum with the peak of the distribution located} above the critical
energy below which triple interactions dominate over the binary ones. Broader
initial distributions, even the Planck spectrum, contain too many photons at
low energies, and triple interactions such as bremsstrahlung quickly eliminate
excess photons, preventing the condensation. This is the reason why the
cooling of photons by electrons proposed by Zeldovich and Levich does not lead
to photon condensation.

Likewise fermion degeneracy manifests itself in relativistic systems such as
white dwarfs and neutron stars, it is possible that BEC of photons might be as
well observed in astrophysical conditions. Given a transient character of
condensation and short time of its existence, the necessary condition is a
supply of soft nonthermal photons, which might provide support for
condensation on longer timescales. A possible candidate could be gamma-ray
bursts, whose time resolved spectra are often fit with a cut-off power law
function, which, as we have shown, is expected when BEC occurs.

Regarding the possibility to observe this phenomenon in the laboratory, we
propose to study the interaction of X-ray lasers with dense plasma targets.
Our example in nonrelativistic case refers to initial conditions with
isotropic distribution of photons with energy about $10$ keV, and number
density about few $10^{28}$ $cm^{-3}$ interacting with Maxwellian electrons
with temperature $50$ keV. The discovery of photon condensation at
relativistic temperatures $k_{B}T_{\gamma}>m_{e}c^{2}$, reported in this work,
makes it possible to think about initial conditions, when dense
non-equilibrium photon gas at MeV energies itself generates electron-positron
pairs and undergoes such a condensation.

\section*{Acknowledgement}

This work is supported within the joint BRFFR-ICRANet-2018 funding programme.


\begin{thebibliography}{99}                                                                                               %
\bibitem {1995Sci...269..198A}M.~H. Anderson, J.~R. Ensher, M.~R. Matthews,
C.~E. Wieman, and E.~A. Cornell, \newblock Science 269, 198 (1995).

\bibitem {1995PhRvL..75.1687B}C.~C. Bradley, C.~A. Sackett, J.~J. Tollett, and
R.~G. Hulet, \newblock Physical Review Letters 75, 1687 (1995).

\bibitem {1995PhRvL..75.3969D}K.~B. Davis et~al., \newblock Physical Review
Letters 75, 3969 (1995).

\bibitem {2002Sci...298..199D}H.~Deng, G.~Weihs, C.~Santori, J.~Bloch, and
Y.~Yamamoto, \newblock Science 298, 199 (2002).

\bibitem {2006Natur.443..430D}S.~O. Demokritov et~al., \newblock Nature 443,
430 (2006).

\bibitem {2010Natur.468..545K}J.~Klaers, J.~Schmitt, F.~Vewinger, and
M.~Weitz, \newblock Nature 468, 545 (2010).

\bibitem {1969JETP...28.1287Z}Y.~B. Zeldovich and E.~V. Levich, \newblock
Soviet Journal of Experimental and Theoretical Physics 28, 1287 (1969).

\bibitem {1924ZPhy...26..178B}S.~Bose, \newblock Zeitschrift fur Physik 26,
178 (1924).

\bibitem {Ein1}A.~Einstein, \newblock Sitzungber. Kgl. Preuss. Akad. Wiss. ,
261 (1924).

\bibitem {Ein2}A.~Einstein, \newblock Sitzungber. Kgl. Preuss. Akad. Wiss. , 3 (1925).

\bibitem {1980stph.book.....L}L.~D. Landau and E.~M. Lifshitz,
\newblock {\em {Statistical physics. Pt.1, Pt.2}}, \newblock Oxford: Pergamon
Press, 1980.

\bibitem {2008bcdg.book.....P}C.~J. Pethick and H.~Smith,
\newblock {\em {Bose-Einstein Condensation in Dilute Gases}}, \newblock
Cambridge University Press, 2008.

\bibitem {2014JPhB...47x3001K}J.~Klaers, \newblock Journal of Physics B Atomic
Molecular Physics 47, 243001 (2014).

\bibitem {1956JETP...04..730K}A.~S. Kompaneets, \newblock Soviet Journal of
Experimental and Theoretical Physics 4, 730 (1956).

\bibitem {1986PhyA..139..165M}E.~E. M\"{u}ller, \newblock Physica A
Statistical Mechanics and its Applications 139, 165 (1986).

\bibitem {1969Ap&SS...4..301Z}Y.~B. Zeldovich and R.~A. Sunyaev,
\newblock\apss {\bf 4}, 301 (1969).

\bibitem {1970Ap&SS...7....3S}R.~A. Sunyaev and Y.~B. Zeldovich,
\newblock\apss {\bf 7}, 3 (1970).

\bibitem {2017PhRvA..95f3611M}J.~T. Mendon\c{c}a and H.~Ter\c{c}as,
\newblock\pra {\bf 95}, 063611 (2017).

\bibitem {2013PhRvA..88a3615K}A.~Kruchkov and Y.~Slyusarenko,
\newblock\pra {\bf 88}, 013615 (2013).

\bibitem {2019arXiv190207998M}P.~Mati, \newblock arXiv:1902.07998 ,
arXiv:1902.07998 (2019).

\bibitem {2007PhRvL..99l5003A}A.~G. Aksenov, R.~Ruffini, and G.~V.
Vereshchagin, \newblock Phys.~Rev.~Lett. 99, 125003 (2007).

\bibitem {2009PhRvD..79d3008A}A.~G. Aksenov, R.~Ruffini, and G.~V.
Vereshchagin, \newblock\prd {\bf 79}, 043008 (2009).

\bibitem {2019PhLA..383..306P}M.~A. Prakapenia, I.~A. Siutsou, and G.~V.
Vereshchagin, \newblock Physics Letters A 383, 306 (2019).

\bibitem {2010PhRvE..81d6401A}A.~G. Aksenov, R.~Ruffini, and G.~V.
Vereshchagin, \newblock\pre {\bf 81}, 046401 (2010).

\bibitem {2017rkt..book.....V}G.~V. Vereshchagin and A.~G. Aksenov,
\newblock {\em {Relativistic Kinetic Theory}}, \newblock Cambridge University
Press, 2017.

\bibitem {Berestetskii1982}V.~B. Berestetskii, E.~M. Lifshitz, and V.~B.
Pitaevskii, \newblock {\em {Quantum Electrodynamics}}, \newblock Elsevier, 1982.

\bibitem {1952RSPSA.215..497M}F.~Mandl and T.~H.~R. Skyrme, \newblock
Proceedings of the Royal Society of London Series A 215, 497 (1952).

\bibitem {2004epb..book.....H}E.~Haug and W.~Nakel,
\newblock {\em {The Elementary Process of Bremsstrahlung}}, \newblock World
Scientific Publishing Co, 2004.

\bibitem {1976spr..book.....J}J.~M. Jauch and F.~Rohrlich,
\newblock {\em The Theory of Photons and Electrons}, \newblock
Springer-Verlag, 1976.

\bibitem {2018JCoPh.373..533P}M.~A. Prakapenia, I.~A. Siutsou, and G.~V.
Vereshchagin, \newblock Journal of Computational Physics 373, 533 (2018).

\bibitem {1976oup..book.....H}G.~Hall and J.~M. Watt,
\newblock {\em {Modern Numerical Methods for Ordinary Differential Equations}},
\newblock New York, Oxford University Press, 1976.

\bibitem {Prakapenia2019}M.~A. Prakapenia and G.~V. Vereshchagin, \newblock(in preparation).
\end{thebibliography}

\end{document}